\documentclass[a4paper]{jpconf}
\usepackage{graphicx}
\usepackage{upgreek}

\begin{document}
\title{The H1 Data Preservation Project}

\author{D M South$^1$ and M Steder$^2$,\\ on behalf of the H1 Collaboration}

\address{Deutsches Elektronen Synchrotron, Notkestra{\ss}e 85, 22607 Hamburg, Germany}

\ead{$^1$david.south@desy.de, $^2$michael.steder@desy.de}

\begin{abstract}
The H1 data preservation project was started in 2009 as part of the global data preservation initiative in
high-energy physics, DPHEP.
In order to retain the full potential for future improvements, the H1 Collaboration aims for level 4 of the
DPHEP recommendations, which requires the full simulation and reconstruction chain as well as the data
to be preserved for future analysis.
A major goal of the H1 project is therefore to provide secure, long-lived and validated access to the H1 data
and analysis software, which is realised in collaboration with DESY-IT using virtualisation techniques.
By implementing such a system, it is hoped that the lifetime of the unique $ep$ collision data from HERA
will be extended, providing the possibility for novel analysis in the future.
The preservation of the data and software is performed alongside a consolidation programme of digital and
non-digital documentation, some of which dates back to the early 1980s.
A new organisational model of the H1 Collaboration, reflecting the change to the long term phase,
is to be adopted in July 2012.
\end{abstract}

\section{Introduction}
\label{sec:intro}

Following the end of data taking in 2007, the H1 Collaboration has developed a data preservation project to
ensure the longevity of the unique $ep$ collisions from HERA.
At the same time, H1 and DESY have become a leading force in the global data preservation
study group DPHEP~\cite{dphep}.
The different data preservation models defined~\cite{dphep_paper} by the DPHEP study group are presented
in table~\ref{tab:levels}.
As new theories or new experimental methods are likely to be the prime reasons for re-analysing the H1 data,
scenarios may arise where only a full and complete preservation model will provide the necessary ingredients.
For example if a cut in the current reconstruction turns out to have been too harsh, or a new simulation model,
written in an alternative computing language, requires a new interface to the existing code.
With this in mind, H1 plans to follow a DPHEP level~$4$ preservation model, details of which are described in
section~\ref{sec:spsystem}.
This level of preservation will necessarily include the full range of both experiment-specific
and external software dependencies, although attempts to minimise the latter are carried out in the
initial step.
However, the benefit of such a model is that the full physics analysis chain is available and full flexibility
is retained for future use.
H1 has also engaged in a multi-faceted documentation programme, described in section~\ref{sec:docu},
which includes employing new methods and means of storing digital and non-digital documentation for the future.
The new organisation model of the H1 Collaboration for the long term phase is described in section~\ref{sec:future}.

\begin{table}[]
  \begin{center}
    \caption{\label{tab:levels}Data preservation levels defined by the DPHEP study group~\cite{dphep_paper}.}
    \centering
    \begin{tabular}{{c}{l}{l}}
      \br
      Level & Preservation Model & Use Case\\
      \br
      1 & Provide additional documentation & Publication related info search \\
      \mr
      2 & Preserve the data in a simplified format & Outreach, simple training analyses\\
      \mr
      3 & Preserve the analysis level software and & Full scientific analyses,\\
      & data format & based on the existing reconstruction\\
      \mr
      4 & Preserve the simulation and reconstruction & Retain the full potential of the\\
      & software as well as basic level data & experimental data\\
      \br
    \end{tabular}
  \end{center}
\end{table}

\section{H1 data samples and software framework}
\label{sec:h1data}

The good and medium quality H1 raw data comprises around $75$~TB and is the basic format to be preserved.
A full set of Compressed Data Storage Tape (CDST) data for the 1996-2007 period comprises about $20$~TB,
and the analysis level files (H1OO, see below) are around $4$~TB.
Other data, such as random trigger streams, noise files, cosmic-data, luminosity-monitor and other calibration
data amounts to only a few TB. Standard Monte Carlo (MC) sets for preservation are also being defined, where the
total data volume is likely to be similar to the real data.
The total preservation volume, including MC and non-collision data, is conservatively estimated to be
about $0.5$~PB.


The H1 reconstruction and simulation software, which creates DSTs from the raw data, is written mainly in
Fortran but also contains some C and C++.
The MC simulation takes the generator files as input and passes them through a GEANT~\cite{geant3} based
simulation of the H1 detector, which takes the relevant run conditions from a database, to produce MC events
in the same format as the data, with some additional information.
The same reconstruction software as used on the data is then applied to the simulated MC events.
%


The majority of physics analysis performed by the H1 Collaboration is done using a common C++
analysis framework, H1OO~\cite{h1oo_chep2010}.
This has had huge benefits in terms of shared analysis code, expert knowledge and calibrations,
working environments and, perhaps most importantly, handling the actual data, where the whole
collaboration uses the same file format, and more often than not the same files.
The H1OO framework is based on ROOT~\cite{root}, and uses its functionality for I/O, data handling,
producing histograms, visualisation and so on.
ROOT also provides attractive solutions for code documentation, which are fully utilised by H1.
Given the level of use in the HEP community, especially at the LHC, it is expected that ROOT will
continue to be supported in the long term.
Major development of the analysis software was essentially completed with the recent $4.0$ release series,
which was developed for and in parallel to the final version of the reconstructed data, DST $7$.

\section{The DESY validation framework}
\label{sec:spsystem}

\begin{figure}[t]
  \begin{center}
    \includegraphics[width=0.85\textwidth]{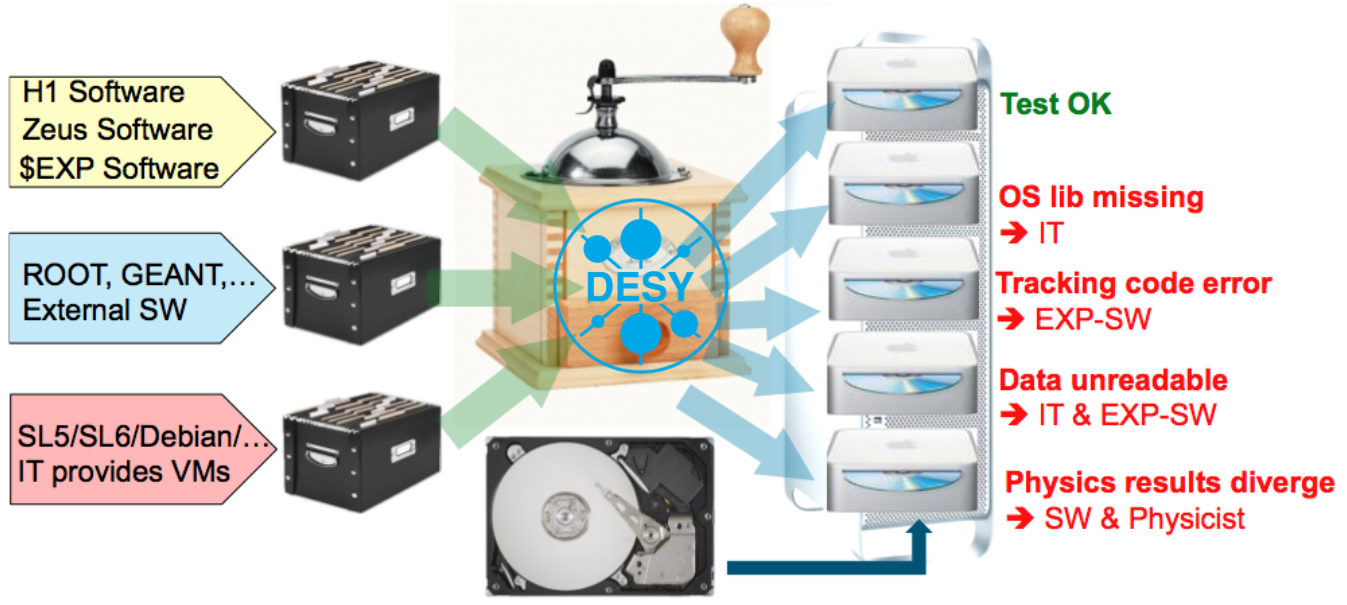}
    \caption{\label{fig:sp-system}An illustration of the idea behind the generic validation framework at DESY.}
  \end{center}
\end{figure}

In order to substantially extend the lifetime of analysis capability it is beneficial to migrate
to the latest software versions and technologies for as long as possible~\cite{yves_chep2012}.
In collaboration with DESY-IT, a framework has been developed to automatically test and validate the
software and data against such changes and upgrades to the environment, as well as changes to the
experiment software itself.
An illustration of this software preservation system, known as the {\tt sp-system}, is given in
figure~\ref{fig:sp-system}.
Technically, this is realised using a virtual environment hosting different configurations of operating
systems (OS), experimental software, and any necessary external dependencies, which are treated as
separate inputs.
In this way, a variety of environments can be provided on one coherent system.
More details on the {\tt sp-system} can be found in~\cite{yves_chep2012}.

\begin{figure}[t]
  \begin{center}
    \includegraphics[width=0.85\textwidth]{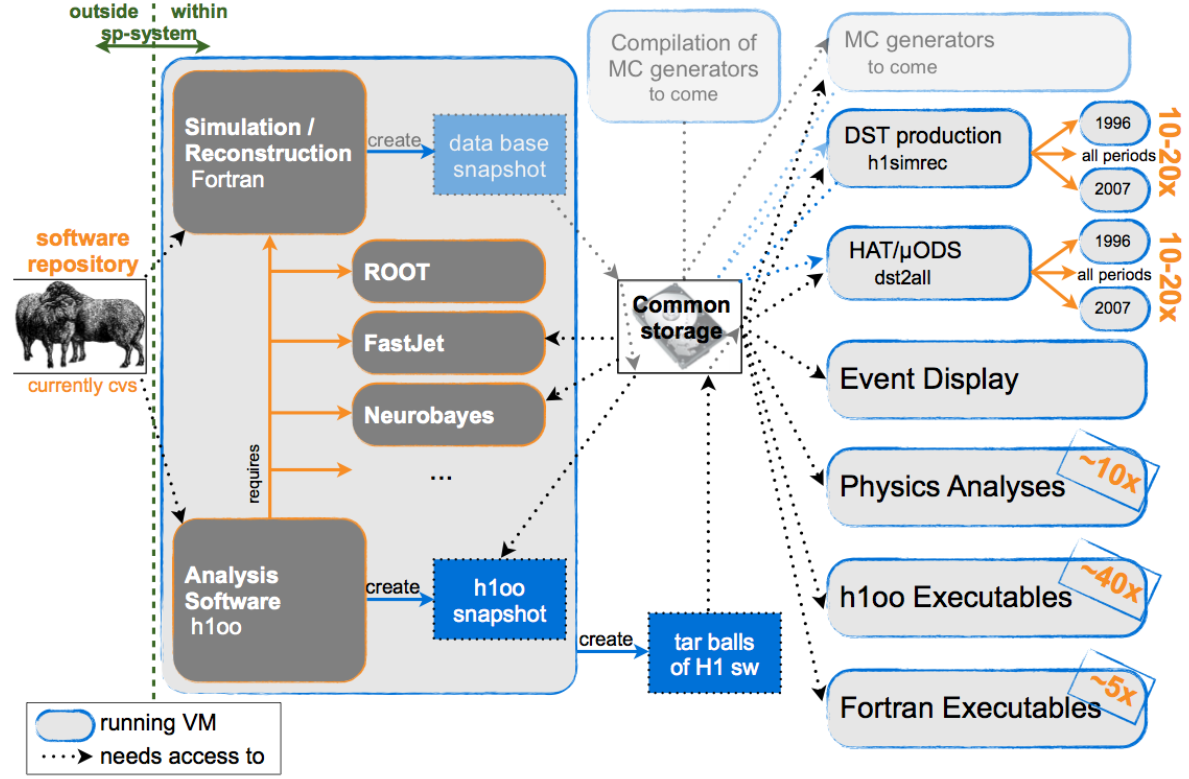}
    \caption{\label{fig:structure}An overview of the preliminary set of tests employed by the H1 Collaboration in
      the DESY validation system.}
  \end{center}
\end{figure}

A crucial component of the {\tt sp-system} is the robost definition of the experimental tests.
As mentioned above, H1 plans a full DPHEP level~$4$ preservation scheme, and the preliminary structure
of the tests to be installed by the H1 experiment to validate the full analysis chain is shown in
figure~\ref{fig:structure}.
The left hand-side details the compilation of experimental and external software.
This is considered as a series of tests, where the compilation of approximately $100$ individual packages
is carried out.
The resulting binaries are stored as tar-balls on a central storage facility within the {\tt sp-system}, where they
are then accessible and used in further tests, described on the right hand side of figure~\ref{fig:structure}.
These tests are wide reaching, examining all areas of the H1 software including among others file production
(``DST production'', ``HAT/$\upmu$ODS''), comparison of analysis histograms (``Physics Analyses'') and
execution of experiment specific tools and macros (``h1oo Executables'', ``Fortran Executables'').
It is estimated that a total of around 250 tests are required to successfully validate the complete analysis chain,
although it should be noted the implementation is still within the development phase.


The main OS used within H1 is Scientific Linux DESY 5 (SLD5), which is fully supported by DESY-IT
with support expected from the distributor until at least 2017.
The recent migration to SLD5, which is the base level for the preservation project, was used to streamline
the software and identify potential future problems.
The current default OS for H1 is SLD5/32-bit and the current focus of the H1 validation effort is on the full migration
to 64-bit operating systems, starting with SLD5/64-bit.
Compilation of all H1 simulation, reconstruction and analysis level software has been successfully achieved
with only minor modifications, where external dependencies (ROOT, Neurobayes~\cite{neurobayes},
FastJet~\cite{fastjet}) are included in the compilation as long as they are not centrally provided.
The inclusion of the compilation of MC generators is ongoing.
Files can now be produced and accessed using libraries and executables created within the {\tt sp-system}.
First checks in the validation system yield promising results and the {\tt sp-system} has allowed SLD5/64-bit
compatibility to be evaluated by the H1 Collaboration. 
As soon as SLD5/64-bit is fully validated, which is expected by the end of 2012, the system will be
used to rigorously test SLD6/64-bit.
As this next generation OS will only be available in 64-bit, this demonstrates the value of the current evaluation. 
Furthermore, the lack of 32-bit support for next generation hardware might make the use of a 64-bit OS
necessary well before 2017.

\begin{figure}[t]
  \begin{center}
    \includegraphics[width=0.85\textwidth]{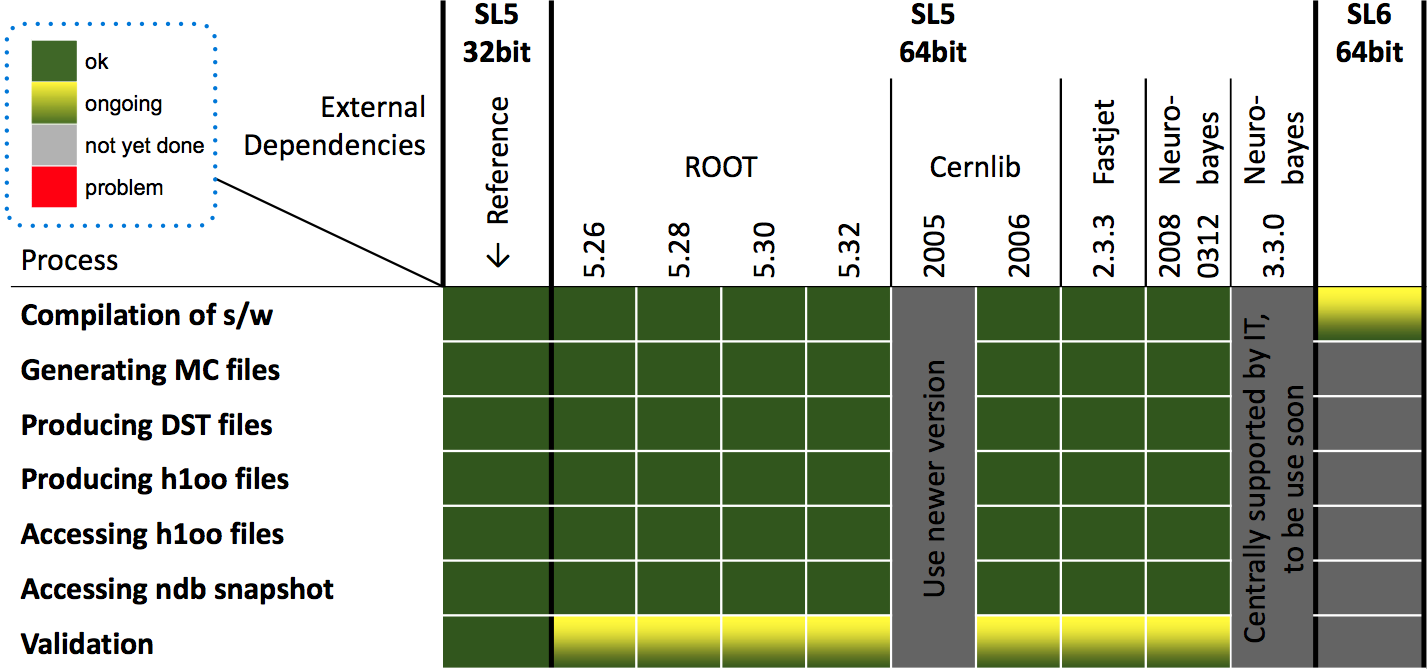}
    \caption{\label{fig:status}Status of the individual steps of the H1 software validation project for
      different operating systems and external dependencies.}
  \end{center}
\end{figure}

Each test-job started in the {\tt sp-system} is assigned a unique ID, and all scripts and input files used in
the test as well as all output files are kept.
This allows validation of all versions against each other and ensures reproducability of previous results.
In addition to this unique ID, all H1 validation jobs are tagged with a description, indicating which software
versions were used, and the Unix timestamp of the execution to aid the book-keeping.
A script-based webpage lists all available runs for a given description and indicates the status of the
compilation for the individual packages with a coloured table cell, which is linked to the corresponding
output file.
The headline cells of this table open another webpage displaying the results of the various validation tests
run for this version.
Again the results are indicated by coloured table cells, which are linked to additional information, like
histograms or ROOT files.
The current status of the H1 validation tests in the {\tt sp-system} is summarised in figure~\ref{fig:status}.

\section{Digital and non-digital documentation}
\label{sec:docu}

A general survey of the state of the non-digital H1 documentation has been performed.
There is a great deal of paper documentation: H1 physics and technical talks from pre-web days;
detector schematics and blueprints; artefacts from the experimental hall such as logbooks.
A future location large enough to store all the documentation for preservation has been secured
in the DESY-Library.
However, the cataloguing and organisation of large quantities of documentation has been a significant
task that could only be done by someone with expert knowledge of the H1 Collaboration.

\begin{figure}[b]
  \begin{center}
    \includegraphics[width=0.85\textwidth]{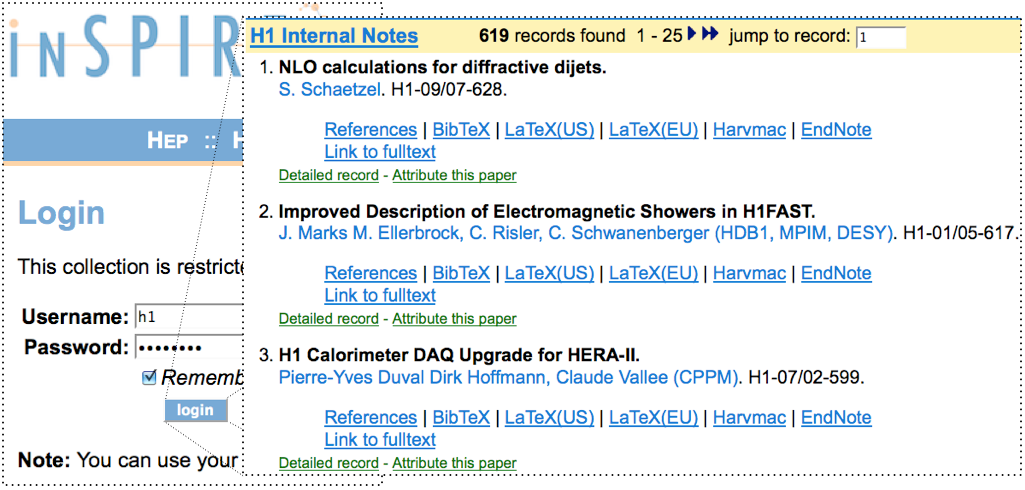}
    \caption{\label{fig:inspire}Screenshot of H1 internal notes on INSPIRE, currently under password protection.}
  \end{center}
\end{figure}

The INSPIRE~\cite{inspire} project has offered, via the DESY-Library, to aid the documentation effort and
several projects are underway with the HERA collaborations, including the ingestion of internal notes, as
illustrated in figure~\ref{fig:inspire}, the ingestion of theses and electronically cataloguing publication
histories: preliminary results, paper and referee reports, versions of paper drafts, and so on.
The large scale digitisation of older H1 documentation is also underway, although due to the volume
of material this has required prioritisation, where preference has been given to plenary meetings.

A great deal of digital H1 documentation exists, mainly but not exclusively on the official webpages.
This includes published papers and preliminary results, review articles and expert notes.
In addition, talks from meetings, conferences, lectures, and university courses are also available.
There are also many unpublished articles, including various notes and the internal wiki-pages that are
extensively used by H1.
Data quality information (physics and technical) and other electronic documentation like H1 software
manuals and notes also contribute.
In-house DDL documentation of Fortran software (h1banks) is being updated.
As mentioned above, the H1OO analysis level software benefits from the automatically generated
HTML documentation from ROOT, but only if the code is correctly written, and any missing information
will be addressed.
Old online shift tools contain much metadata and are particularly vulnerable to loss.
Such information has in the main not been updated since July 2007 and electronic logbooks
(shift, trigger and other detector components) and detailed run information contained in the system
supervisor, which was used to log the collision data as it was taken, is to be secured.
The recent migration of the main H1 webserver to a DESY-IT supported VM resource will help secure
the H1 digital documentation for the future.

\section{The new operational model of the H1 Collaboration}
\label{sec:future}

\begin{figure}[t]
  \begin{center}
    \includegraphics[width=0.8\textwidth]{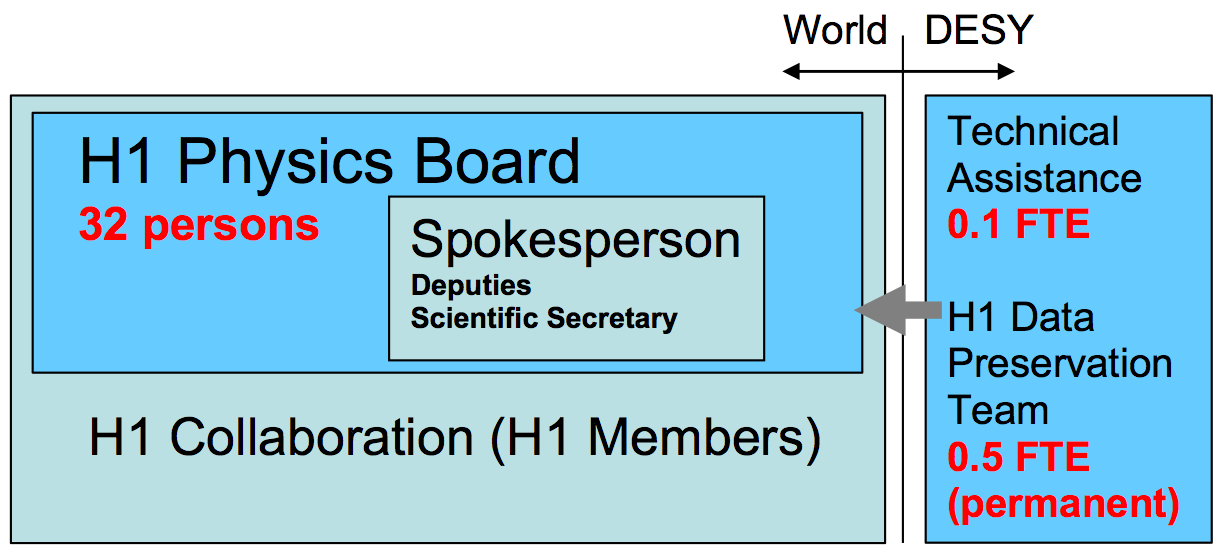}
    \caption{\label{FutureH1}The new organisational model of the H1 Collaboration, employed from July 2012.}
  \end{center}
\end{figure}

A new model for long-term governance of the H1 Collaboration was formerly adopted in 2011,
with the transition to the new model taking place in July 2012.
In the new model, the H1 Collaboration Board (H1CB), which includes representatives from all participating
institutes and the H1 Executive Committee, which is a smaller structure elected by the H1CB to meet more
regularly, are both replaced by the H1 Physics Board (H1PB).
The mandate of this board, which comprises a broad selection of H1 members from all physics and
technical working groups is: to be the general contact point for H1 physics and data beyond the
collaboration lifetime; to communicate with the host lab, DESY, and other experiments; to supervise
the H1 data; to maintain contact with the global DPHEP initiative; and to overview potential
future publications using H1 data.
The new organisational model is illustrated in figure~\ref{FutureH1}.

\section{Conclusions}
\label{sec:conc}

The H1 Collaboration has made significant progress in the data preservation projects begun in the last few years.
The validation of the full H1 software environment using up to date operating systems is now possible using
the validation framework developed at DESY.
The development of a full and robust set of validation tests is now underway to safeguard H1 analysis
for the future.
At the same time, a comprehensive evaluation of the state of H1 documentation is progressing well, including
new initatives with INSPIRE, in which DESY leads the global effort.
The H1 Collaboration will move to a new operational model in July 2012.

\section*{References}

\end{document}